\documentclass[10pt,conference,final,letterpaper]{IEEEtran}

\ifCLASSINFOpdf
  \usepackage[pdftex]{graphicx}
  \graphicspath{{../figures/}}
  \DeclareGraphicsExtensions{.pdf}
\else

\fi

\ifCLASSINFOpdf
 	\usepackage[pdftex]{graphicx}
\else
\fi

\ifCLASSOPTIONcompsoc
  \usepackage[caption=false,font=normalsize,labelfont=sf,textfont=sf]{subfig}
\else
  \usepackage[caption=false,font=footnotesize]{subfig}
\fi

\usepackage{amsmath}
\usepackage{graphicx}
\usepackage{caption} 
\usepackage{multirow}
\usepackage[ampersand]{easylist}
\usepackage{amssymb}
\usepackage{amsthm}
\usepackage{amsmath}
\usepackage{hyperref}
\usepackage[euler]{textgreek}
\usepackage{url}
\usepackage[table]{xcolor}

\begin{document}

\title{Toward Evaluating the Complexity to Operate a Network}

\author{
	\IEEEauthorblockN{
		\textbf{Marc Bruyere}	
}
	\IEEEauthorblockA{
 		IIJ Innovation Institute \\ 
Japanese-french laboratory for informatics (CNRS IRL 3527) \\
The University of Tokyo\\
Japan
	}
		\IEEEauthorblockN{
		\textbf{Christoff Visser}	
}
	\IEEEauthorblockA{
 		IIJ Innovation Institute \\ 
 		Japan
	}
\and
\IEEEauthorblockN{
 		\textbf{Daphne Tuncer}
}
	\IEEEauthorblockA{
	Imperial College London \\
 		United Kingdom
	}
}

\maketitle

\begin{abstract}
The task of determining which network architectures provide the best ratio in terms of operation and management efforts \textit{vs.} performance guarantees is not trivial. In this paper, we investigate the complexity of operating different types of architectures from the perspective of the space of network parameters that need to be monitored and configured. We present OPLEX, a novel framework based on the analysis of YANG data models of network implementations that enables operators to compare architecture options based on the dimension of the parameter space. We implement OPLEX as part of an operator-friendly tool that can be used to determine the space associated with an architecture in an automatic and flexible way. The benefits of the proposed framework are demonstrated in the use case of Internet Exchange Point (IXP) network architectures, for which we take advantage of the rich set of publicly available data. We also exploit the results of a survey and direct consultations we conducted with operators and vendors of IXPs on their perception of complexity when operating different architectures. OPLEX is flexible, builds upon data models with widespread usage in the community, and provides a practical solution geared towards operators for characterizing the complexity of network architecture options.
\end{abstract}

\IEEEpeerreviewmaketitle

\section{Introduction}
\label{sect:Intro}

Operators usually have the choice between different architecture options to implement their network. These options typically respond to two main needs: 1) deploying what is essential in terms of functionality to support the offered services, and 2) ensuring that the network can evolve in the face of future changes and/or unexpected events. 

From the perspective of an operator, the choice of architecture has an impact on the efforts required to operate and manage the network. The type of architecture not only affects the time it takes to make changes and fix issues, it also drives the expertise and knowledge needed to perform network management tasks. Having mechanisms in place to evaluate the complexity of operating a network based on its architecture can thus support the operator achieve a better understanding of future expected efforts. 

In the past, different approaches were proposed to quantify the complexity associated with the operation and management of a network \cite{clemmCisco07} \cite{bensonNSDI09}\cite{sunCONEXT13}. A key contribution of previous work is the definition of quantifiable metrics enabling systematic comparison across network implementations \cite{whiteBook15}\cite{RFC7980}. In general, the factors of complexity of operating a network are based on three dimensions \cite{ behringerRAI09}: \textit{i)} the operator, represented by his/her expertise; \textit{ii)} the management interface, based on its degree of abstraction and automation; and \textit{iii)} the network, described by the space of its configuration parameters. 

In this paper, we revisit the link between evaluating the complexity of operating a network and characterizing the network through its parameter space, in the light of recent developments in the community towards the implementation of standardized models of network functionality. More specifically, we present OPLEX (Operation comPLEXity), a novel framework based on the analysis of YANG \cite{rfc7950} models of network architectures to determine and compare architecture options based on the dimension of the space of parameters that an operator needs to monitor and configure in order to manage underlying resources and services. In contrast to previous solutions, \textit{e.g.,} \cite{bensonNSDI09}\cite{sunCONEXT13}, OPLEX is agnostic to an operator's internal standard specifics. Instead it builds upon a standardized format with widespread usage across vendors and operators \cite{yangCatalogue}. In addition, it takes into account the whole functionality space of a network as enabled through the analysis of YANG models available for the full network stack at the device level, \textit{i.e.,} from optical transport to routing policies and management (see OpenConfig \cite{openconfigDM} for example). 

Using OPLEX we develop a flexible tool that can automatically extract the dimension of the network parameter space associated with a network architecture. Our tool is generic, \textit{i.e.,} it can accommodate the YANG data models of any vendor implementation. It is also easily extensible, \textit{i.e.,} new data models can easily be added to enrich the set of implementations for which complexity scores can be computed. 

To illustrate the functionality of OPLEX, we elaborate on the complexity of operating a network in the specific use case of Internet eXchange Point (IXP) network architectures, for which we collected qualitative and quantitative datasets. In addition to the importance IXPs have in today's in the Internet ecosystem \cite{agerIXP12}\cite{bruyereJSAC18}, another motivation for focusing on this type of network comes from the fact that IXPs have traditionally been engaged in an open approach to their business, as well as to their technical and performance specifications. Information about IXPs are publicly available from different sources on the web (\textit{e.g.,} \cite{peeringDB}\cite{pch}\cite{IXPDB}), and individual webpages. 

Based on OPLEX tool, we compare the parameter space of network architectures used in today’s IXPs.  We discuss the implication on network complexity by putting the obtained results in perspective with a qualitative dataset of key operational considerations for the operators of these networks that we collected by disseminating a survey within the IXP community and engaging in direct consultations with multiple actors in this domain. Our results reveal how IXP operators perceive the complexity and knowledge required to design and maintain nine selected architectures that represent up-to-date solutions to interconnect Internet autonomous networks. 

The remainder of this paper is organized as follows. We provide background information on evaluating the complexity of operating a network and assessing the space of network parameters in Sect. \ref{sect:background}. We then present the details of the OPLEX framework in Sect. \ref{sect:OPLEXframework}, and describe its implementation as a tool in Sect. \ref{sect:OPLEXTool}. We elaborate on the Internet Exchange use case in Sect. \ref{sect:ixp} where we introduce the datasets used in this paper. Finally, we discuss how OPLEX contributes to the state-of-the art in Sect. \ref{sect:relatedWork}, and provide concluding remarks and directions for our future work in Sect. \ref{sect:conclusions}.

\section{Complexity and Network Parameter Space}
\label{sect:background}

\begin{figure}[t]
\centering
\includegraphics[trim = 0cm 1cm 0cm 3.5cm, clip, scale=0.26]{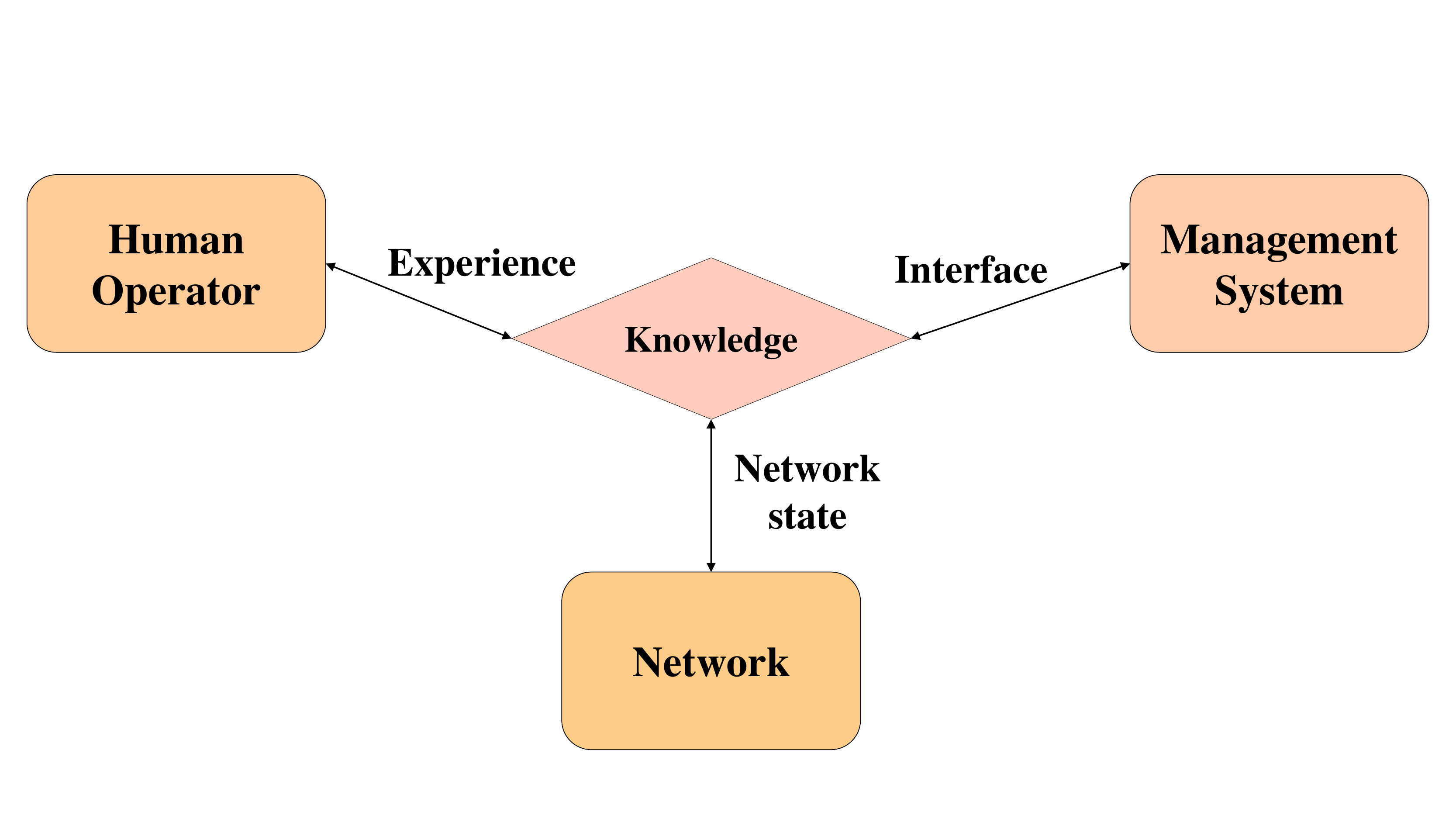}  
\caption{Knowledge as a connector between the dimensions of the complexity of operating a network.}
\label{fig:knowledge}
\end{figure}

Various factors contribute to the complexity of operating a network. In a seminal paper \cite{behringerRAI09} Behringer proposes to model factors of complexity based on three dimensions: the network, the management system and the (human) operator. As depicted in Fig. \ref{fig:knowledge}, \textit{knowledge} acts as a connector between the three dimensions. More specifically, to operate the network, the operator builds upon knowledge of the environment, which gets enriched through experience. This knowledge can be described in terms of network state, \textit{i.e.,} a set of parameters representing both the hardware and software resources that need to be configured and/or monitored. Read and write access to that state is mediated through the management system that provides an interface between the network and the operator. The complexity of operating a network is a translation of that knowledge. It can be apprehended from three perspectives:
\begin{enumerate}
	\item by measuring the experience and level of expertise of the operator.
	\item by evaluating the degree of sophistication of the management interface, \textit{i.e.,} support for abstraction, automation and intelligence.
	\item by quantifying the dimension of the network parameter space. 
\end{enumerate}

\subsection{Network Functional Domains}
\label{sect:functionalDomains}

Evaluating the complexity an operator faces operating and managing different types of network architectures through the dimension of the network parameter space is the underlying principle of most previously proposed approaches \cite{bensonNSDI09}\cite{sunCONEXT13}\cite{brownIM05}\cite{clemmCisco07}. It comes from the observation that the larger the parameter space, the harder it is to maintain a full knowledge of that space, and hence to track, identify and correct operational issues \cite{clemmCisco07}. 

In a networking context, the parameter space covers multiple functional domains, ranging from the physical infrastructure to the provided services. Functional domains can be classified in four main groups.  
\begin{itemize}
	\item Facility: all physical/virtual resources deployed for hosting and powering the network infrastructure.
	\item Interconnection: all physical/virtual resources and processes to enable local and remote connectivity across the network.
	\item Communication: all physical/virtual resources and processes to enable information exchange across the network.
	\item Services: all physical/virtual resources and processes to provide added value services on top of the infrastructure (\textit{e.g.,} security) 
\end{itemize}

Each group involves a variety of functional elements, including networking functions and equipment, for which different design and implementation choices can be selected, \textit{e.g.,} layer-2 switching mechanisms, network operating system, switching platforms, \textit{etc.} The combination of these functional elements define the parameter space of a network, and by extension, drives how complex it is to operate that network. 

A key challenge to evaluate the complexity is to determine the granularity at which to take functional elements into account so as to be representative of the network operations. For instance, a naive approach would consist in assessing complexity as a count of the number of activated networking functions (\textit{e.g.,} EVPN, MPLS, port filtering, \textit{etc.}). Such an approach is oblivious to device and function state whereas this state is critical to network operations. In line with previous initiatives \cite{bensonNSDI09}, we define in this work the functional elements at the granularity of network configurations. This approach offers a good trade-off between practicality, \textit{i.e.,} it is easy for the operator to extract, and expressiveness, \textit{i.e.,} it includes all parameters taken into account to reason upon network operations. As opposed to previous work, however, we investigate the use of standardized network data models.  

\subsection{Normalized Network Parameter Space}
\label{sect:normalizedMeasure}

A main objective when evaluating the complexity of operating a network is to enable comparison between different network architectures. To provide a fair ground for comparison, it is essential to have a reference point. This requires analyzing all relevant data model files, from which the network parameter space can be extracted. By nature, the extracted space depends on the input data models. In order to provide a generic method for evaluating the complexity - and hence enable comparison between architectures based on their associated parameter space - it is essential for the extraction to rely on normalized procedures, which can be realized by using standardized data models. 

Various standards exist for modeling network state and configurations (\textit{e.g.,} Structure of Management Information, Managed Object Format, \textit{etc.}). In this work, we focus on the YANG data modeling language \cite{rfc7950}. YANG models build upon a recognized standard in the industry, with major vendors supporting YANG releases of their implementations. They are also used by several standard bodies, \textit{i.e.,} IETF, IEEE, ETSI. The description provided by YANG models reflects the specifics of an implementation. In the last few years, the OpenConfig organization has been working towards the development of a set of vendor-neutral YANG data models based on a generic abstraction of networking elements (functions, services and protocols) \cite{openconfigDM}. The process of releasing OpenConfig implementations is currently ongoing. Current models are grouped based on 34 categories \cite{openconfigGit} that range the traditional networking layers. Today's major vendors offer an OpenConfig-integrated version of YANG models of their implementations (see \cite{openconfigGit}). A recent release by Juniper presents a didactic mapping of Juniper device commands to OpenConfig syntax \cite{juniperOpenConf}. The availability of such a rich source of data, consolidated around a generic abstraction of networking functionality, makes YANG the ideal candidate for developing a methodology to determine and analyze the space of parameters associated with various network architectures.

\section{OPLEX Framework}
\label{sect:OPLEXframework}

The objective of OPLEX is to determine the space of state and configuration parameters associated with a network architecture by analyzing relevant YANG \cite{rfc7950} data models. In this section we start by providing a quick walk-through YANG models, and further discuss the methodology we developed to extract the parameter space and determine its dimension. 

\subsection{YANG Model}
\label{sect:ymodelOverview}

YANG \cite{rfc7950} is a standardized data modelling language for network management protocols. It provides modeling primitives for network device state (read only parameter such as packet counter), device configurations (read/write parameters such as interface name, addresses, \textit{etc.}), remote procedure calls and notifications. While originally developed as a data model for the Network Configuration Protocol (NETCONF) \cite{rfc6241}, it can support other management agents, \textit{e.g.,} RESTCONF \cite{rfc8040}. 

YANG organizes data definitions into hierarchies of schema nodes, \textit{i.e.,} data structures of parameter definitions and attributes, grouped into modules. Each module constitutes a self-contained object that can be compiled. Modules fall under two categories \cite{rfc8199}: \textit{1)} network element modules that provide data definitions for device-centric functions (\textit{e.g.,} IPv4, Ethernet); and \textit{2)} network service modules that describe network-level services (\textit{e.g.,} L2VPN, VPLS). As a language, YANG follows a set of syntactic rules and conventions that facilitate extensibility (\textit{e.g.,} \texttt{augment} statement) and reusability (\textit{e.g.,} \texttt{import} statement). 

\subsection{Methodology}
\label{sect:ymodelAnalysis}

To determine the set of state and configuration parameters relevant to a network architecture, OPLEX analyzes YANG models by proceeding at three levels of representation:
\begin{itemize}
	\item module level by extracting state and configuration data.
	\item the device level by selecting modules corresponding to the functions used to achieve network operations, \textit{e.g.,} (switching protocol, redundancy mechanism, link aggregation feature, \textit{etc.}) 
	\item the network instance level by analyzing the characteristics of the instantiation of the network architecture (\textit{e.g.,} connectivity, activated interfaces, \textit{etc.}).
\end{itemize} 

\subsubsection{Module analysis}

Information about state and configuration data is contained in two types of nodes: \texttt{leaf} and \texttt{leaf-list}. \texttt{leaf} nodes are representations of state or configuration parameters which they model through an identifier and a data type; for instance \texttt{leaf interface\_name type string}. \texttt{leaf-list} nodes are sequences of leaf nodes of a particular type. In a similar fashion to \texttt{leaf}, they come with an identifier, \textit{e.g.,} \texttt{leaf-list vlan-id type string}. YANG enables multiple instances of \texttt{leaf} and \texttt{leaf-list} to be declared by defining them as child nodes of specific constructs called \texttt{list} nodes. \texttt{list} constructs are used to define an interior data node in the hierarchy of schema nodes. Each \texttt{list} can be the child node of another \texttt{list} node, forming as such dependency structures (example: \texttt{list destination-group (list destination (list config))}). Extracting these dependencies is essential as they contribute to the dimension of the parameter space. OPLEX thus determines both the set of all \texttt{leaf} and \texttt{leaf-list} defined in a YANG module, as well as the associated \texttt{list} dependencies if relevant. 

\subsubsection{Device analysis}

In an operational context a only subset of the functions implemented in a device is actively employed. Examples of functions include for instance the type of protocol used to route traffic, the link aggregation feature selected to increase link capacity or the type of mechanisms triggered to provide redundancy guarantees. OPLEX determines all YANG modules relevant to active device-level elements of a specific network architecture implementation. Selecting the set of appropriate modules is however challenging given that parameters associated with an active function can be defined in more than one module. In this paper we achieve the selection through lexicographic matching between the conventional name of the protocols / technologies related to activated functions and the name of the YANG modules. 



\subsubsection{Network instance analysis}

\texttt{leaf}, \texttt{leaf-list} and \texttt{list} extracted from individual YANG modules define the network parameter space of a specific implementation. While the dimension of that space does not depend on the actual value of these parameters, it is affected by the size of \texttt{leaf-list} and \texttt{list}, which is driven by the instantiation of that specific implementation (for instance the number of configured interfaces depends on the number of devices and their connectivity). OPLEX consolidates the set of state and configuration parameters by extracting from network instance characteristics \texttt{leaf-list} and \texttt{list} size information.   

\subsection{Parameter Space Dimension}
\label{sect:complexityScore}
 
We define the dimension of the parameter space of a network architecture implementation as a count on the total number of parameters that can be accessed via \texttt{read} and \texttt{write} operations. 

Let $\mathcal{M}_{d}$ be the set of YANG modules relevant to the functions activated on a device $d$. In addition let $\mathcal{I}_{m}$ be the set of \texttt{leaf} and $\mathcal{J}_{m}$ the set of \texttt{leaf-list} extracted from module $m$. We denote as $|j|$ the number of elements in \texttt{leaf-list} $j \in \mathcal{J}$. We also denote as $\mathcal{L}_{i}$ and $\mathcal{L}_{j}$ the set of lists in the list dependencies associated with \texttt{leaf} $i \in \mathcal{I}$ and \texttt{leaf-list} $j \in \mathcal{J}$, respectively. Let $|l_{i}|$ be the number of elements in list $l_{i} \in \mathcal{L}_{i}$ and $|l_{j}|$ be the number of elements in list $l_{j} \in \mathcal{L}_{j}$. The dimension $\delta_{d}$ associated with device $d$ is equal to:

\[
\delta_{d} = \sum_{m \in \mathcal{M}_{d}} \bigg( \sum_{i \in \mathcal{I}_{m}} u_{i} + \sum_{j \in \mathcal{J}_{m}} v_{j}  \cdot |j| \bigg)
\]

with $u_{i}$ a variable equal to $1$ if $\mathcal{L}_{i} = \emptyset$ and to $\prod_{l_{i} \in \mathcal{L}_{i}} |l_{i}|$ otherwise, and $v_{j}$ a variable equal to $1$ if $\mathcal{L}_{j} = \emptyset$ and to $\prod_{l_{j} \in \mathcal{L}_{j}} |l_{j}|$ otherwise.\\

The dimension can be determined in a flexible way and be computed at the module level:

\[
\forall m \in \mathcal{M}, \quad \delta_{m} = \sum_{i \in \mathcal{I}_{m}} u_{i} + \sum_{j \in \mathcal{J}_{m}} v_{j}  \cdot |j|
\]

It can also be agnostic to the specifics of either or both environment and activated functions. In this case, all $|j|$, $|l_{i}|$ and $|l_{i}|$ take a default value of 1 (environment-agnostic) and $\mathcal{M}_{d}$ includes all modules defined in device $d$ (function-agnostic): 

\[
\delta_{agnostic} = \sum_{m \in \mathcal{M}_{d}} \bigg( I_{m} + J_{m} \bigg)
\] 

with $I_{m}$ the size of set $\mathcal{I}_{m}$ and $J_{m}$ the size of set $\mathcal{J}_{m}$.

\subsection{Discussion}
\label{sect:frameworkDiscussion}

The development of OPLEX comes with a number of considerations. As explained in Section \ref{sect:ymodelOverview}, being able to select all modules relevant to a specific active function is a challenging process. Our current selector employs lexicographic matching based on naming convention used in networking, which in some cases may be too coarse to determine the relevance of a module. As a result, the dimension of the parameter space may be over-estimated. While we plan in the future to investigate more fine-grained approaches for implementing the selection process (by creating a dictionary of names from branches in YANG models, for instance), we note here that being able to determine the exact value of the dimension of the parameter space is not discriminating. Given that the objective of OPLEX is to compare network architectures, what it is essential is to obtain comparative values for $\delta$, which can be achieved by using a reference point in terms of device model. We discuss this aspect in more detail in Section \ref{sect:toolIllustration}. 

The objective of determining the space of network parameters associated with a network architecture is to apprehend the complexity to operate that type of architecture through the network factor, as described in Section \ref{sect:background}. OPLEX focuses on the count of parameters at the device level. It does not take into account the relationship that exists between devices in a network, which also contributes to the complexity an operator faces when operating the network. The impact of interactions between devices on complexity was studied by Chun \textit{et al.} in \cite{chunNSDI08}. In the future, we will investigate how to integrate OPLEX with the model of device interactions proposed in \cite{chunNSDI08} to develop a quantifiable measure of complexity to operate a network. 

Finally, without additional information about the instantiation of a network architecture in an operational environment, it is not possible for OPLEX to determine the size of \texttt{list} and \texttt{leaf-list}. As both \texttt{list} and \texttt{leaf-list} contribute to the definition of the parameter space, and hence need to be counted, we set their default value to one. The resulting $\delta$ value can be used as a lower bound on the dimension of the extracted space. 

\section{OPLEX Implementation}
\label{sect:OPLEXTool}

We implement the functionality of the OPLEX framework as part of a tool that can be used to automatically determine the dimension of the parameter space at the device level (as per the definition in Section \ref{sect:complexityScore}) of any input network architecture. Our tool is designed to be operator-friendly: \textit{i)} it is generic to any implementation for which YANG data models are available; \textit{ii)} it is easily extensible, \textit{i.e.,} the repository of module data can be enriched at any time as the YANG models of different implementations become available; \textit{iii)} it is practical, \textit{i.e.,} the user can specify vendor, function and network characteristics information and select the options to determine the dimension, whether at the device or module level, or if it is agnostic.    

\subsection{Tool Functionality}

\begin{figure}[t]
\centering
\includegraphics[trim = 0cm 0cm 0cm 0cm, clip, scale=0.45]{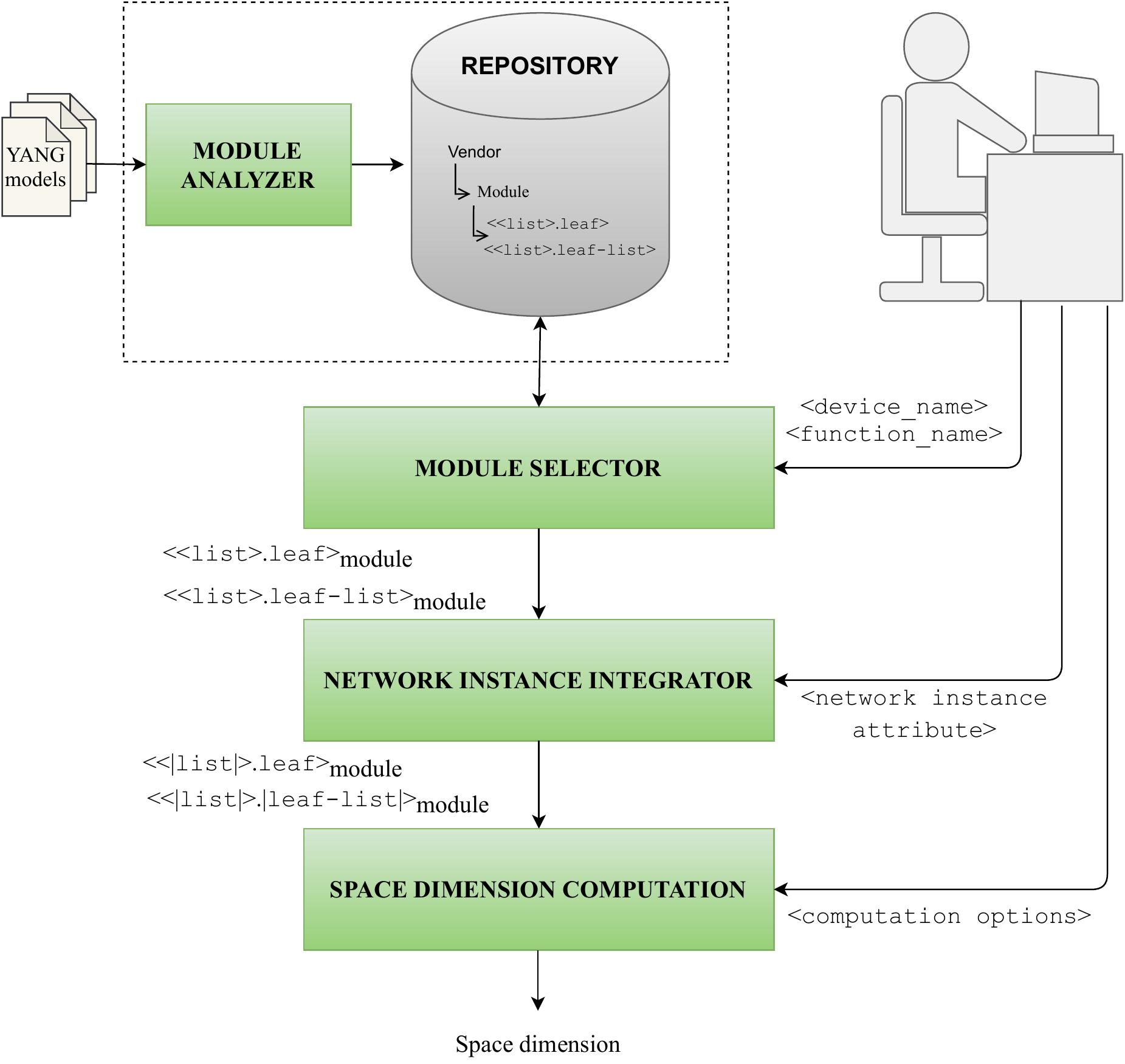}  
\caption{Overview of the process to determine the dimension of the parameter space of an input architecture.}
\label{fig:scoringProc}
\end{figure}

The main components of the OPLEX tool are depicted in Fig. \ref{fig:scoringProc}. It includes four main functions: 1) module analyzer, 2) module selector, 3) network instance integrator, and 4) space dimension computation. 

\textbf{Module analyzer} This component is responsible for extracting all \texttt{leaf}, \texttt{leaf-list} and \texttt{list} parameters from an input set of YANG modules. Parameters are stored in a global repository where they are organized per vendor and per module. In our current implementation of the tool, the extraction follows a two-step process. Input YANG modules are first converted into their tree diagram representation \cite{rfc8340} using the tree output format of the YANG transformer tool \textit{pyang} \cite{pyang}. The obtained tree structures are then parsed to extract the set of \texttt{leaf}, \texttt{leaf-list}, and associated \texttt{list} parameters related to each module. This two-step process enables us to verify that the input modules are free from formatting and syntactic errors\footnote{We came across formatting and syntactic issues for some specific implementations. These were reported to the interested parties.}. We plan to integrate this verification as part of a single-step process in future development of the tool.    

\textbf{Module selector} This component is responsible for selecting in the repository the set of \texttt{leaf}, \texttt{leaf-list} and associated \texttt{list} of all active functions related to a network architecture. Information about the activated functions and the device model (\textit{i.e.,} vendor name) are provided as input either directly by the user, \textit{e.g.,} the operator, or by interfacing the tool to an external (management) system. Modules relevant to the input device model and active functions are selected by performing a lexicographical match against vendor names and module names, respectively, from data contained in the repository. In case no functions and/or no device names are provided in input, the module selector retrieves by default parameter information for all modules of an OpenConfig device representation. 

\textbf{Network instance integrator} This component is responsible for determining \texttt{leaf-list} and \texttt{list} size information of the parameters associated with the set of modules selected by the module selector based on network instance characteristics provided as input in terms of attributes. These attributes concern connectivity, port and interface settings, in particular. In the absence of information regarding the network instance characteristics, the network instance integrator assigns a default value of $1$ to the size of all \texttt{leaf-list} and \texttt{list}.  

\textbf{Space dimension computation} This component implements the functions described in Section \ref{sect:complexityScore} and computes the relevant dimension(s) based on inputs received from the network instance integrator. Computation options (\textit{i.e.,} type of function $\delta$) can also be specified by the user. By default, the score is calculated for an OpenConfig device with all functions activated. 

\subsection{Illustration}
\label{sect:toolIllustration}

We use OPLEX to determine the dimension of the parameter space extracted from three examples of network device families for which YANG models are publicly available: the Cisco Nexus series \cite{ciscoYANG}, the Juniper junOS series \cite{juniperYANG} and OpenConfig \cite{openconfigGit}. These examples are illustrative of different levels of network abstraction, and hence different definitions of parameter space. OpenConfig YANG models are based on a high-level vendor-neutral definition of device parameters \cite{openconfigDM}. Cisco YANG models builds upon OpenConfig representation but extends definitions with vendor features. Juniper YANG models provide vendor-specific fine-grained definitions of the network parameters. In each case we compute the value $\delta_{agnostic}$ by taking into account all available modules.

The obtained values of $\delta_{agnostic}$ are in line with the degree of abstraction provided by each network device family. OpenConfig provides higher level abstraction of parameter definition compared to the two vendors and obtained a value of $\delta_{agnostic}$ equal to $4,620$. By building on top of OpenConfig, the definition of the parameters in Cisco devices is made at an intermediate level of abstraction compared to Juniper, which translates into a value of $\delta_{agnostic}$ one order of magnitude lower, ranging from $21.10^{4}$ to $26.10^{4}$ for the Cisco Nexus series (on a total of 18 devices) and from $10^{5}$ to $40.10^{5}$ for the Juniper junOS series (on a total of 75 devices). These results show that having a reference point in the form of a model of network functionality is essential to compare architectures. In the next section, we use the YANG models of OpenConfig as a reference to compare architectures in the specific use case of IXP networks.

\section{IXP Use Case}
\label{sect:ixp}

IXPs constitute the core public infrastructure of the Internet. Their main service is to provide layer-2 connectivity to BGP autonomous systems. To be called an IXP, at least three Internet operators need to be connected through the same peering Local Area Network (LAN). Internet Service Providers and Content Delivery Networks exchange Internet traffic through the IXP. IXPs not only reduce the portion of the traffic that an ISP delivered via its upstream transit providers, they also reduce latency and increase security. 

In this paper we focus on the use case of IXPs for three main reasons. First, IXP architecture and operations are well documented. Information is available from a rich set of publicly available data sources, which enables us to conduct an evaluation of the impact of architecture options based on real data. In addition, given that all IXPs share a common goal and core service, it is possible to compare network architectures based on the knowledge required to operate alternative options. Finally, we take advantage of our long-term involvement within the IXP community to collect a dataset of qualitative results regarding the perception of IXP operators with respect to their experience operating different types of architectures.   

\subsection{IXP Overview}

The scale of IXP infrastructures is diverse. It ranges from small networks with a single switch to large infrastructures involving tens of nodes. To determine the distribution of today's IXP infrastructure sizes, we analyze information about $841$ IXPs available from PeeringDB \cite{IXPDB}. PeeringDB provides details about the links connecting operators and the peering LAN, and it is updated directly by all the Internet actors \cite{Lodhi14}. 
To extract IXP sizes, we proceed in two steps. We first filter out from all $841$ entries those associated with a valid IXP, where an entry is deemed valid if it satisfies three criteria: \textit{i)} there is at least one facility\footnote{A facility in PeeringDB is a point of presence}, \textit{ii)} there are at least three active operators connected, and \textit{iii)} there are no active links with null speed. We obtain $533$ valid IXPs. For each valid IXP, we then determine the number of operated switches. Given that this value is not directly provided in PeeringDB, we develop a simple method to estimate it. We assume that all switches have 48 ports (typical top of the rack switch) with eight ports allocated to up-links or other purposes. We then count at least one switch per facility and one more for every 40 ports per location. 

\begin{figure}[hbt!]
\centering
\includegraphics[scale=0.55]{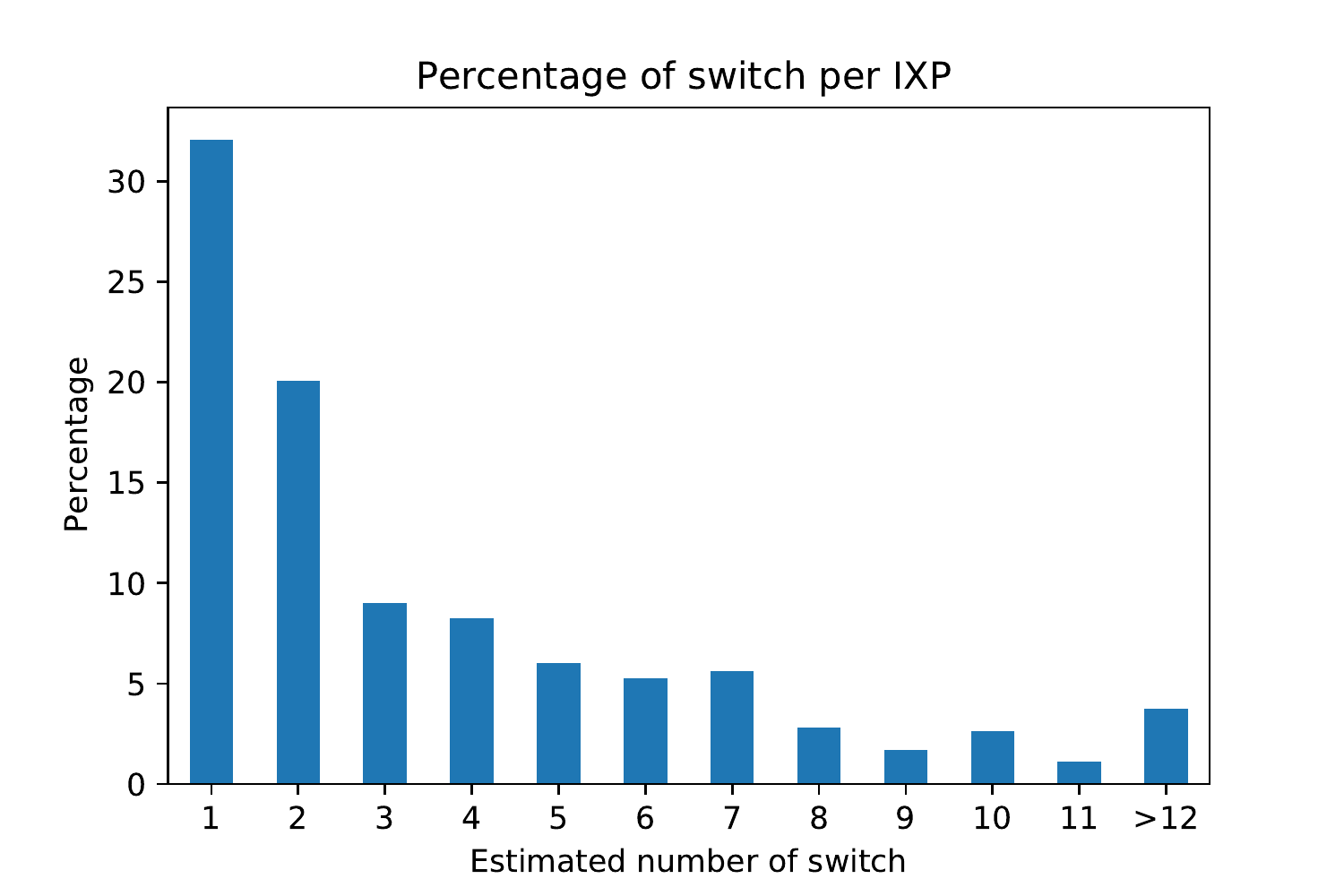}  
\caption{Estimated number of switches per IXP.}
\label{fig:num_SW_per_IXP}
\end{figure}

The distribution of the estimated number of switches per IXP for the $533$ valid IXPs is shown in Fig. \ref{fig:num_SW_per_IXP}. As can be observed, more than 30\% operate a single switch only. In fact the majority of these IXPs have an infrastructure of less than 3 switches (53\%).  

\subsection{IXP Architectures and Parameter Space}

To deliver connectivity services, IXPs can use different architectures, usually as a function of the size of their infrastructure. We investigate typical options deployed by IXPs today based on the document released by EURO-IX \cite{EURO-IX_Wish-List} that provides the list of features and architectures expected from network vendors. EURO-IX is the most significant association of IXPs, involving the largest players, \textit{i.e.,} DE-CIX, AMS-IX, and LINX. To consolidate the list of common architectures, we also directly consult IXPs website and relevant press-releases (in particular given that the wish-list was published in 2013\footnote{It is worth highlighting that architecture changes are not rapid processes for IXPs.}). We finally validated the extracted list with three major network vendors used by IXPs.    

The main architectures and protocols are shown on the first two rows of Table \ref{tab:IXP_Score}. Although the main service of an IXP is to provide layer-2 connectivity to Internet operators, it can be noted that in order to scale, their service can be implemented on a layer-3 overlay architecture. In general Link Aggregation (LAG) and Spanning Tree protocol implementations are used by small and medium IXPs, \textit{i.e.,} with less than four switches, while larger IXPs rely on layer-3 VxLAN or layer-3 overlay type of networks.

We use our OPLEX tool to extract the parameter space of each of the architectures presented in Table \ref{tab:IXP_Score} and determine its dimension. The results are reported in the last two rows of the table. The row before the last indicates the value of $\delta_{agnostic}$ at the device-level for the associated architecture. The last row presents the OPLEX score that we define as the ratio between the value $\delta_{agnostic}$ of the relevant architecture to the value $\delta_{agnostic}$ of the most straightforward layer-2 (LAG) architecture (complexity score $2,386$ used as baseline). The highest relative complexity score ($1.79$) is obtained for the most complex layer-3 overlay architecture implementing ISIS, MPLS and BGP protocols.

\begin{table*}[!t]
 \label{table:IXP_archi_score}
\centering{
\begin{tabular}{|l|c|c|c|c|c|c|c|c|c|}
\hline
Architecture & \multicolumn{2}{c|}{\cellcolor{blue!20}Layer2 Only} & \multicolumn{3}{c|}{\cellcolor{blue!40}Layer3 VxLAN } & \multicolumn{4}{c|}{\cellcolor{blue!60}Layer3 Overlay} \\ \hline
Protocols & \cellcolor{blue!20} LAG & \cellcolor{blue!20} STP & \cellcolor{blue!40}Static &\cellcolor{blue!40}IS-IS & \cellcolor{blue!40}OSPF & \cellcolor{blue!60} ISIS & \cellcolor{blue!60} OSPF & \cellcolor{blue!60}ISIS-BGP & \cellcolor{blue!60} OSPF-BGP \\ \hline
OpenConfig-aft & \cellcolor{black!50} & \cellcolor{black!50} & \cellcolor{black!50} & \cellcolor{black!50} & \cellcolor{black!50} & \cellcolor{black!50} & \cellcolor{black!50} & \cellcolor{black!50} &  \cellcolor{black!50}  \\ \hline
OpenConfig-bfd &           &           &   \cellcolor{black!50}     &   \cellcolor{black!50}     &   \cellcolor{black!50}     &  \cellcolor{black!50}    &   \cellcolor{black!50}   &  \cellcolor{black!50}    &  \cellcolor{black!50}    \\ \hline
OpenConfig-bgp &           &           &       &       &      &     &     &  \cellcolor{black!50}   & \cellcolor{black!50}    \\ \hline
OpenConfig-interfaces  & \cellcolor{black!50} & \cellcolor{black!50} & \cellcolor{black!50} & \cellcolor{black!50} & \cellcolor{black!50} & \cellcolor{black!50} & \cellcolor{black!50} & \cellcolor{black!50} &  \cellcolor{black!50}  \\ \hline
OpenConfig-isis &           &           &       &    \cellcolor{black!50}   &       &   \cellcolor{black!50}  &     &  \cellcolor{black!50}   &     \\ \hline
OpenConfig-lacp & \cellcolor{black!50} & \cellcolor{black!50} & \cellcolor{black!50} & \cellcolor{black!50} & \cellcolor{black!50} & \cellcolor{black!50} & \cellcolor{black!50} & \cellcolor{black!50} &  \cellcolor{black!50}  \\ \hline
OpenConfig-local-routing  &           &           &    \cellcolor{black!50}   &   \cellcolor{black!50}    &    \cellcolor{black!50}   &  \cellcolor{black!50}   &  \cellcolor{black!50}   &  \cellcolor{black!50}   &  \cellcolor{black!50}   \\ \hline
OpenConfig-mpls &           &           &       &       &       &  \cellcolor{black!50}   &   \cellcolor{black!50}  &  \cellcolor{black!50}   &  \cellcolor{black!50}   \\ \hline
OpenConfig-network-instance & \cellcolor{black!50} & \cellcolor{black!50} & \cellcolor{black!50} & \cellcolor{black!50} & \cellcolor{black!50} & \cellcolor{black!50} & \cellcolor{black!50} & \cellcolor{black!50} &  \cellcolor{black!50}  \\ \hline
OpenConfig-ospf &           &           &       &       &    \cellcolor{black!50}   &     &  \cellcolor{black!50}   &     &  \cellcolor{black!50}   \\ \hline
OpenConfig-platform & \cellcolor{black!50} & \cellcolor{black!50} & \cellcolor{black!50} & \cellcolor{black!50} & \cellcolor{black!50} & \cellcolor{black!50} & \cellcolor{black!50} & \cellcolor{black!50} &  \cellcolor{black!50}  \\ \hline
OpenConfig-routing-policy &           &           &    \cellcolor{black!50}   &   \cellcolor{black!50}    &    \cellcolor{black!50}   &  \cellcolor{black!50}   &  \cellcolor{black!50}   &  \cellcolor{black!50}   &  \cellcolor{black!50}   \\ \hline
OpenConfig-stp &           &    \cellcolor{black!50}       &       &       &       &     &     &     &     \\ \hline
OpenConfig-system  & \cellcolor{black!50} & \cellcolor{black!50} & \cellcolor{black!50} & \cellcolor{black!50} & \cellcolor{black!50} & \cellcolor{black!50} & \cellcolor{black!50} & \cellcolor{black!50} &  \cellcolor{black!50}  \\ \hline
OpenConfig-terminal-device  & \cellcolor{black!50} & \cellcolor{black!50} & \cellcolor{black!50} & \cellcolor{black!50} & \cellcolor{black!50} & \cellcolor{black!50} & \cellcolor{black!50} & \cellcolor{black!50} &  \cellcolor{black!50}  \\ \hline
OpenConfig-vlan & \cellcolor{black!50} & \cellcolor{black!50} & \cellcolor{black!50} & \cellcolor{black!50} & \cellcolor{black!50} & \cellcolor{black!50} & \cellcolor{black!50} & \cellcolor{black!50} &  \cellcolor{black!50}  \\ \hline
\textbf \textdelta agnostic & 2386  & 2504 & 2684 & 3202  &  2912 & 3499  & 3209 & 4274 & 3984 \\ \hline
\cellcolor{yellow!10} \textbf{OPLEX score}  & \cellcolor{red!0} 1.00 & \cellcolor{red!05} 1.05 & \cellcolor{red!12} 1.12& \cellcolor{red!34} 1.34& \cellcolor{red!22} 1.22  &  \cellcolor{red!47} 1.47 & \cellcolor{red!35} 1.35 & \cellcolor{red!79} 1.79 & \cellcolor{red!67} 1.67 \\ \hline
\end{tabular}
}
\caption{IXP architectures and network parameter space dimension.}
\label{tab:IXP_Score}
\end{table*}

\subsection{Operator Survey and Consultations}

\begin{figure}[hbt!]
\centering
\includegraphics[scale=0.33]{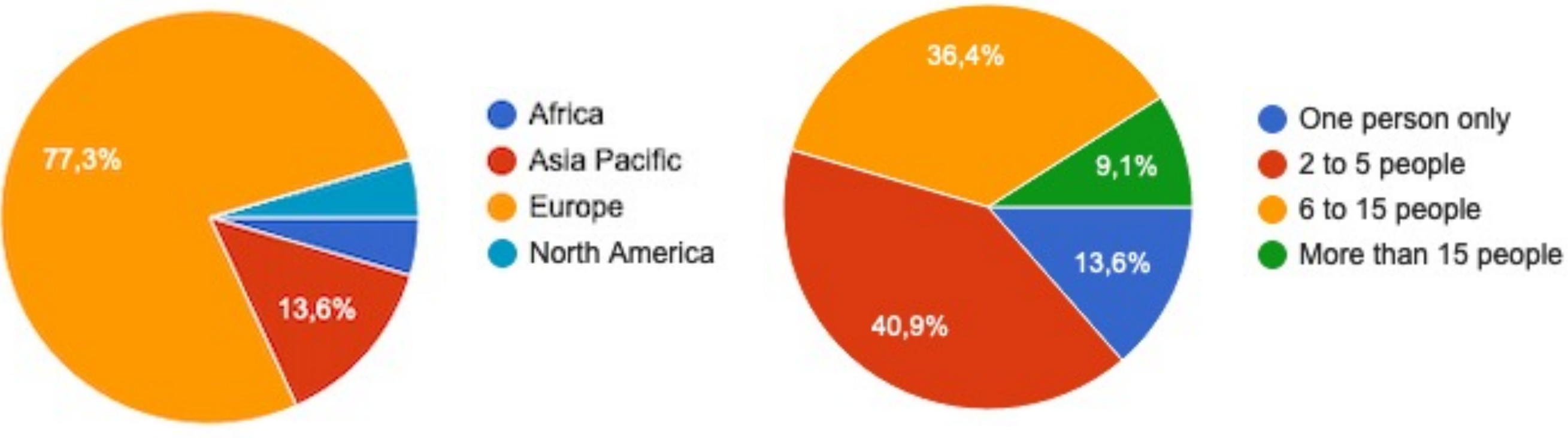}  
\caption{Survey participants geographical origin and size}
\label{fig:pies_chars_survey}
\end{figure}

To understand how operators themselves perceive the efforts they need to provide and manage various types of architectures, we conducted a survey, and further direct consultations, among members of the IXP community. Eighteen IXPs and three major vendors of the IXPs market participated in the survey. Fig. \ref{fig:pies_chars_survey} shows the geographical origin of the respondents, as well as their technical staff size that gives an idea of the involved \textit{human resources}. Most respondents are from Europe, which is consistent with the European concentration of IXPs \cite{ixmap}. We consider our panel as being representative of the IXPs community.

We asked two main questions in the survey: 1) which level of certification is estimated to be necessary to operate different types of architectures, and 2) what perception of complexity - on a scale from 1 to 10 - the operator has with respect to these different architectures. The first question is used to understand the experience and knowledge required to operate and design a network, while the second is intended to put the notion of complexity into perspective from the point of view of the operator.  

\begin{figure}[hbt!]
\centering
\includegraphics[scale=0.36]{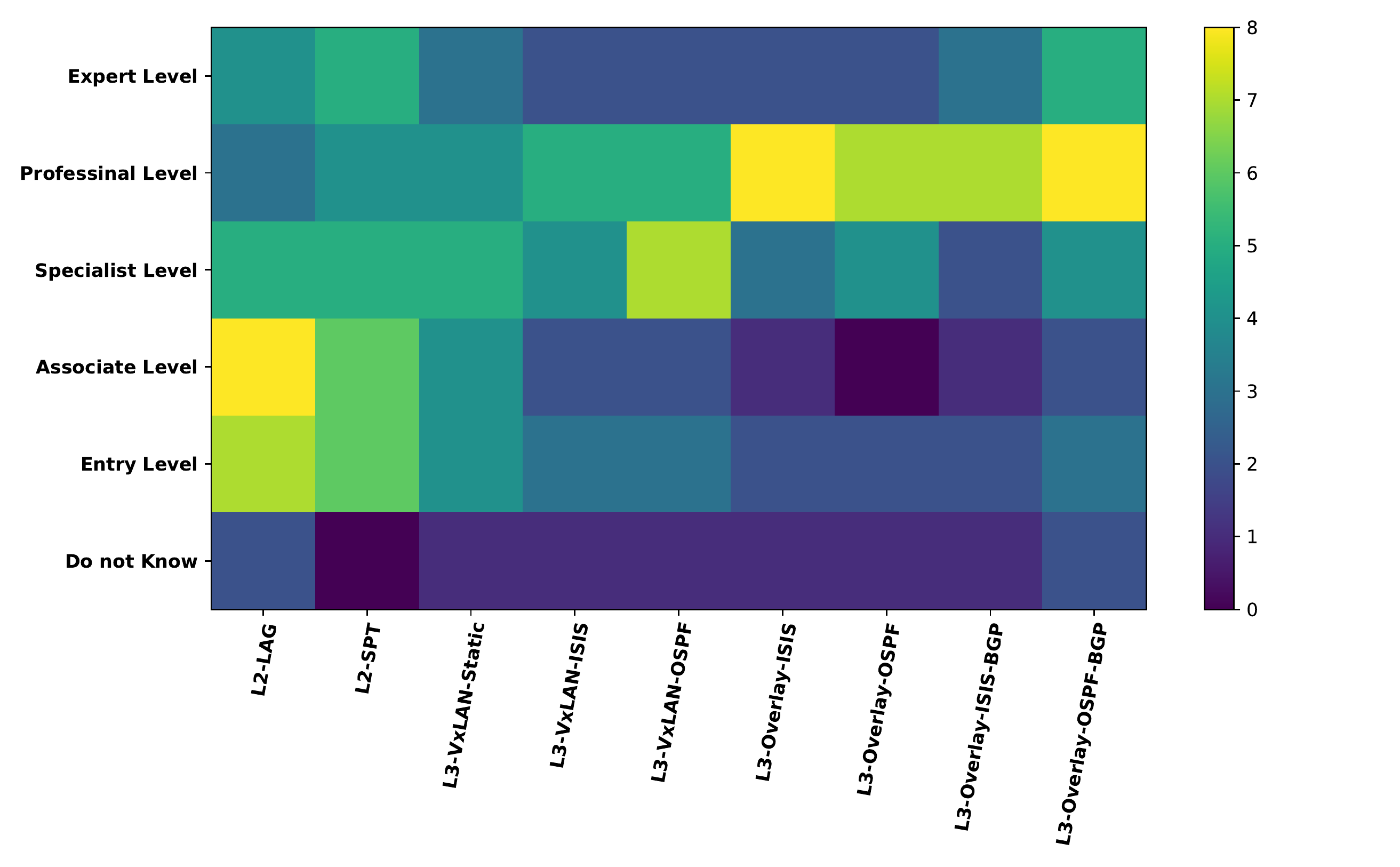}  
\caption{Vendor certification evaluation}
\label{fig:certification_level_survey}
\end{figure}

We used the networking industry training programs validated through certification exams as a knowledge scale for the first question. Certifications include four levels: entry, associate, specialist, and professional. The associated exams are all based on multiple-choice questionnaires and are usually taking place in a certification exam center. For the expert level, the exam also consists in configuring and implementing in a limited-time, state-of-the-art architectures on actual equipment at vendors' certification centers. Fig. \ref{fig:certification_level_survey} shows the responses to the first question in the form of a heatmap. For L2-LAG, most respondents estimate that an associate level is expected. For other architectures, the results show that the expectations of respondents in terms of certification is more diverse, except for L3-Overlay architectures for which most respondents indicate that a professional-level is needed.

\begin{figure}[hbt!]
\centering
\includegraphics[scale=0.44]{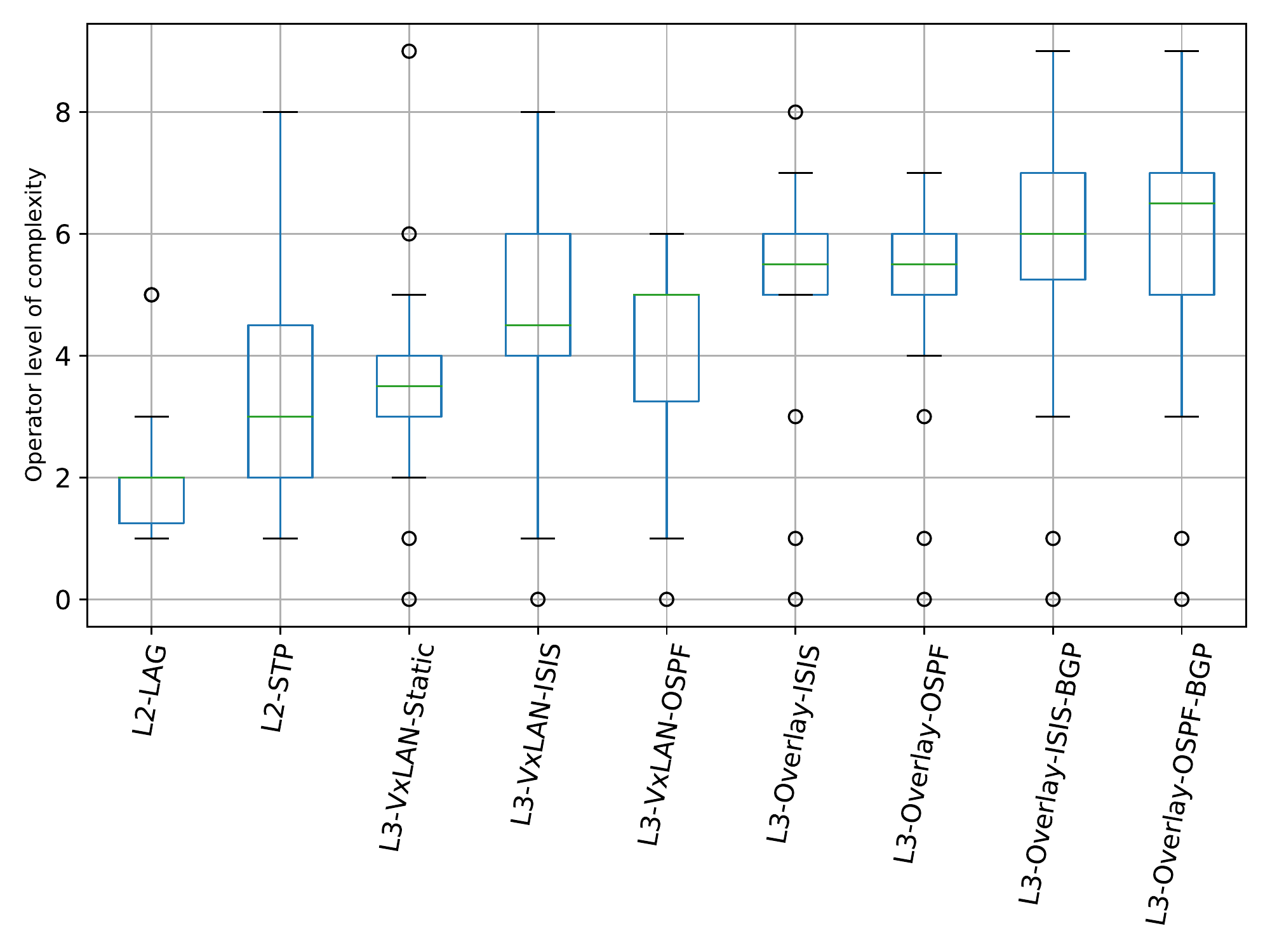}  
\caption{IXP operators complexity level perception per architecture.}
\label{fig:complexity_level_survey}
\end{figure}

The responses to the second question are shown as box plots in Fig. \ref{fig:complexity_level_survey}. The value 0 corresponds to the answer ``I don't know".  The L2-LAG architecture is rated on average at 2. L3-Overlay architectures are usually rated as being two to three times more complex. 

\begin{figure}[hbt!]
\centering
\includegraphics[scale=0.46]{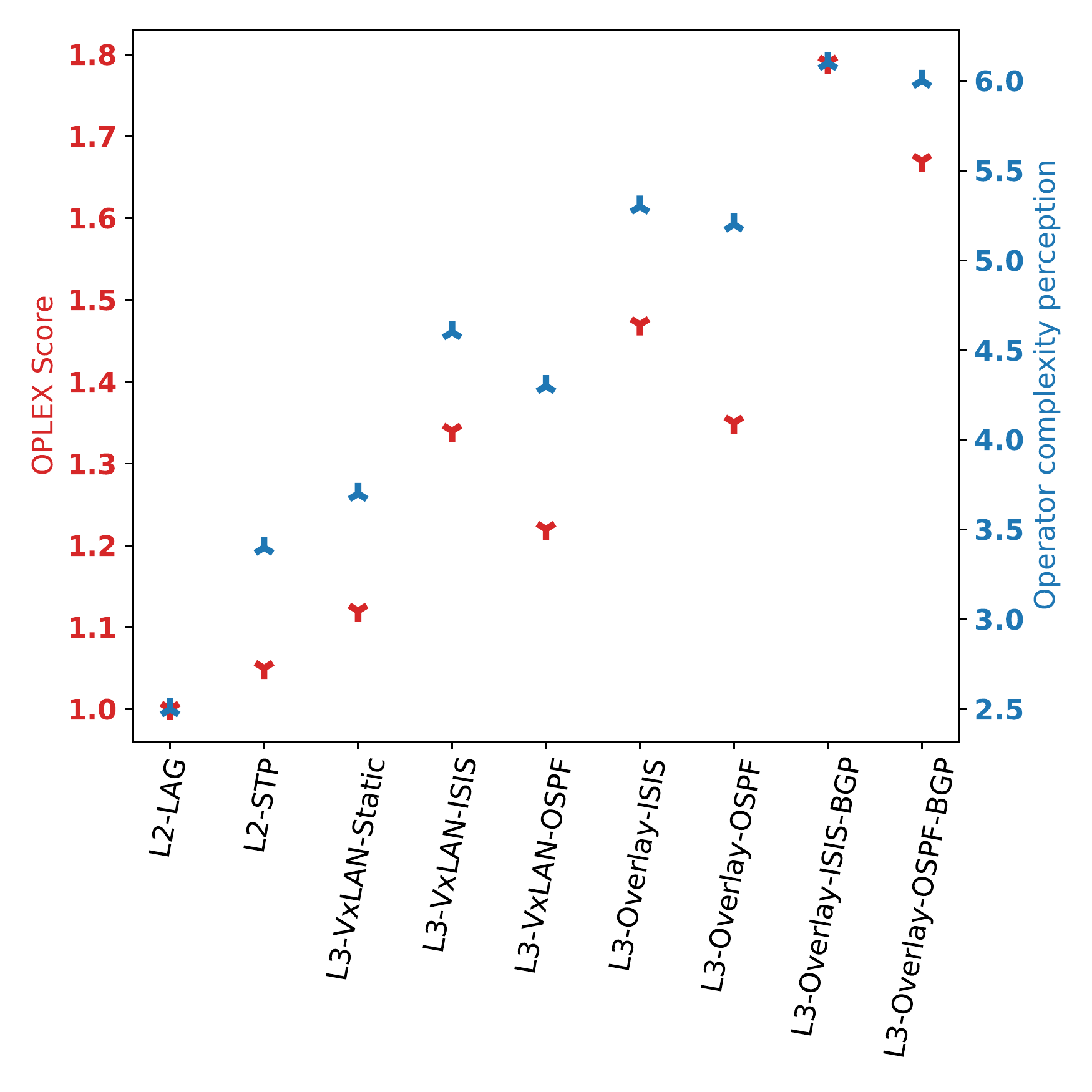}  
\caption{Comparing OPLEX score with the operator perception}
\label{fig:opflex_and_survery_complexity}
\end{figure}

To put the values reported in Table \ref{tab:IXP_Score} in perspective with the perception of complexity as reported by IXP operators who responded to our survey, we superpose the two sets of results in a double y-axis figure in Fig. \ref{fig:opflex_and_survery_complexity}, with OPLEX scores on the left and operator complexity perception on the right. We can see that the size of the parameter space associated with an architecture and the perception of complexity in operating that architecture both increases in the correlated fashion. 

\subsection{Use-Case Discussion}

IXPs, like the rest of the Internet, encounter massive traffic growth. To sustain increasing traffic volume, large IXPs tend to move to layer-3 overlay architectures, which comes with increasing complexity from the perspective of operating the network as shown with the obtained results. Increasing traffic volume also means more devices to manage, which also contribute to increasing the complexity. In order to scale, IXPs face multiple complexity dimensions. OPLEX can help with automated complexity evaluation tools combined at the design phase and network management automation tools to identify the more suitable architecture and protocols stack to keep operations and management as simple as possible.

Another path to simplification can come from programmable data-plane and SDN solutions dedicated to IXPs specific requirements. A key challenge for IXPs is how to operate systems with continuously increasing the complexity at the management level while making sure that desirable levels of performance are maintained. We leave addressing this question for future work.

\section{Related Work}
\label{sect:relatedWork}


The work presented in this paper comes within the literature of research efforts that have been focusing on developing methods for measuring the management and operational complexity of networked communication systems. Relevant approaches include the work by Brown \textit{et al.} \cite{brownIM05}\cite{brownTNSM07}, with subsequent contributions by Clemm \cite{clemmCisco07}, that argue in favor of the definition of operator-facing metrics to quantify the complexity associated with managing network infrastructures. It also encompasses the work by Sch\"onw\"alder \cite{schonwalderIM05} that proposes metrics to analyze the characteristics of Management Information Base (MIB) modules and that evaluates the usage of different features of the data models used in these modules MIBs. In addition, it includes the efforts initiated by Ratnasamy in \cite{ratnasamyHotNets06} and further extended by Chun \textit{et al.} in \cite{chunNSDI08} that focus on the development of a conceptual framework for measuring the complexity of routing protocol implementations. Finally, it covers the proposals presented by Benson \textit{et al.} in \cite{bensonNSDI09} and by Sun \textit{et al.} in \cite{sunCONEXT12} and \cite{sunCONEXT13} which both depend on the analysis of network configuration files to determine a measure of operational network complexity. 

Our work is also motivated from developments in the software engineering domain where solutions for evaluating and managing configuration complexity span theoretical frameworks \cite{krobFormal06}, practical measurement tools \cite{meinickeSoftEng16} and qualitative methodologies and best practices \cite{beyer18}.

By design, OPLEX builds on top of all these proposed solutions to evaluate the complexity of operating a network from the perspective of its configurations. It does however address important limitations of previous work by providing a solution that is easy for operators to use, independent of the specifics of internal configuration standards, and adapted to any type of network architectures for which YANG data models are available. In that respect OPLEX contributes to the efforts engaged in the recent years by White et al. \cite{RFC7980}\cite{whiteBook15} towards formalizing the concept of complexity for the design, deployment, maintenance and management of communication networks and computer networked systems.

\section{Conclusions}
\label{sect:conclusions}

Understanding how complex it is to operate and manage a network is a key challenge for operators. Network architectures that are too complex from an operational point of view can not only be too costly to manage in terms of time and required expertise, they can also impact performance and robustness.

In this paper, we investigate the link between evaluating the complexity to operate a network and characterizing the network through its parameter space using IXPs as a case study. We develop OPLEX, a framework that enables the extraction of the space of state and configuration parameters associated with a network architecture based on the analysis of YANG data models relevant to that architecture. As opposed to previous work, OPLEX takes advantage of standardized data and neutral vendor models to characterize a network architecture. OPLEX is operator-friendly and easy to use, generic to any implementation with available YANG data models, and can be easily extended as more models become available in the community. 

In the future, we plan to extend OPLEX with more vendors and more functions, and take into account additional network characteristics, in order to evaluate other types of architectures. In addition, we plan to integrate the OPLEX framework as part of a general methodology to measure the complexity of operating a network. In general with our work, we aim at participating in the community-driven efforts for the deployment at larger scales of resource abstractions for networks, \textit{e.g.,} OpenConfig, by demonstrating how such realizations can also benefit the implementation of novel ways by which communication network infrastructures can be characterized.   

\section*{Acknowledgement}
The work of Daphne Tuncer is supported by the Imperial College Research Fellowship Scheme. 

We wish to acknowledge the IXPs operators, specially from the EURO-IX community and the networks vendors, for participating in the survey.

\end{document}